\documentclass[pra,showpacs,preprint]{revtex4-1}
\usepackage{graphicx}
\usepackage[usenames]{color}
\usepackage{dcolumn}
\usepackage{amsmath}
\usepackage{amssymb}
\usepackage{bm}

\begin{document}
\input{epsf}
\preprint{APS/123-QED}
\title{CN molecule collisions with H$^+$ at a wide range of astrophysical energies}
\author{Madina R. Sultanova$^1$\footnote{msultanova5@gmail.com},
and Dennis Guster$^1$\footnote{dcguster@stcloudstate.edu}}
\affiliation{Department of Information Systems and BCRL, St. Cloud State 
University,  %
St. Cloud, MN 56301-4498, USA}
\date{\today}    
\begin{abstract}
We analyze the quantum-mechanical rotational excitation/de-excitation spectrum and cross sections
of CN molecules during low and high-energy collisions with protons, H$^+$. The problem is of
significant importance in astrophysics of the early Universe, specifically connected with the problems of cosmic
microwave background (CMB) radiation.
A quantum-mechanical close-coupling method is applied in this work.
The cyanide molecule (CN) is treated as a
rigid rotor, i.e. the distance between the carbon and nitrogen atoms is fixed at an average
equilibrium value. The new results of the excitation/de-excitation cross-sections
and corresponding thermal rate coefficients are compared with the results of few previous calculations
performed on the basis of few approximate semiclassical frameworks. The interaction potential between 
CN and H$^+$ is taken in the following form: proton induced polarization potential + proton-dipole potential +
proton-quadrupole potential.
\end{abstract}

\maketitle

\section{INTRODUCTION}

Matter emits electromagnetic radiation when it has a temperature above zero Kelvin. This is also called thermal radiation because it is generated by thermal motion of particles in the matter. Electromagnetic radiation can be absorbed by matter as well and matter that absorbs all radiation is called a blackbody.  A blackbody has a uniform temperature and produces radiation characterized by the frequency distribution that depends on its temperature. Though a background radiation, referred to as excess noise, was noticed as early as 1941, it wasn't until 1965 that Arno Penzias and Robert Wilson \cite{meyer} used highly sensitive equipment and careful observation, which led to the labeling of this radiation as cosmic microwave background (CMB) radiation: a nearly isotropic (even in all directions) thermal radiation in the Universe. It is know that when a molecule absorbs a photon it moves from a ground state to an excited energy state. Molecules in the space between Earth and distant stars produce absorption lines seen on Earth that correspond to the spectra of those stars. Most of these lines are viewed at the ground state of the molecules, however, researchers found that CN molecules observed in various parts in space all have faint absorption lines at the first excited energy state \cite{field}. Questions arose as to what could excite the CN molecules in space? The temperature of radiation can be found from the amount of molecules excited by that radiation. Measurements of the spectrum of this radiation showed that it was almost identical to blackbody radiation with the temperature of about 3K \cite{takayanagi}. Since these measurements where observed for almost all CN molecules in different parts of the Universe, it showed that whatever was exciting the molecule was distributed fairly evenly throughout space. This radiation cannot be attributed to energy from stars or intergalactic gas, for example, since this radiation's spectrum does not look like the spectrum for stars or intergalactic gas or etc. And also, if the energy of this radiation across the Universe is to be summed up, it would be much too great to be radiation from stars or such. Therefore, researchers believe that whatever is exciting the CN molecules in space ,with a spectrum of a blackbody at 3K, is CMB. 

As mentioned previously, CMB is isotropic, however, not perfectly so. It has an anisotropy of about 0.1 percent, which is due to a Doppler shift caused by the motion of the solar system through CMB. This anisotropy is very small, therefore, CMB is still regarded as almost perfectly uniform radiation in the Universe. One might wonder what caused CMB radiation to occur in the Universe at all? To answer this question, researchers have focused on the blackbody characteristic of the radiation, since this is an intrinsic characteristic. Knowing the present temperature of the radiation being at about 3K, one can calculate the temperature at an earlier time, which turns out to be quite large. This large temperature greatly points to the existence of the the Big Bang: the rapid expansion of the once a dense and compact Universe. Another model that goes together with the Big Bang theory is the model of the expanding Universe. It is strongly shown experimentally and supported theoretically that the Universe is expanding. In the 1920's Edward Hubble found that spectra of distant galaxies are redshifted, which means the galaxies are traveling away from Earth.The model of the expanding Universe states that the cold Universe of today must have had very high temperature and much higher density when it first begun billions of years ago. With temperatures greater than 10$^8$ K, the Universe was once a plasma at thermodynamical equilibrium and made of electrons, protons, and helium ions, that radiated and absorbed photons. As the Universe expanded, the temperature of the plasma and radiation decreased. At temperatures of about 4000K, recombination of protons and electrons occurred and the plasma became a mixture of neutral hydrogen and helium atoms. The Universe became transparent to radiation, as it is now. As the Universe continued to expand, the temperature of the radiation declined but the blackbody character of the radiation remained as a  ``relic." CMB radiation is presumed to be ``relic" radiation from the early times of the Universe.

\section{Quantum-Mechanical Approach}

\subsection{Equations and Methods}

A quantum mechanical method was used in deriving our results for rotational quantum transitions in
collisions between the CN molecule  and a proton $H^+$, that is:
\begin{equation}
\mbox{CN}(j) + \mbox{H}^+ \rightarrow \mbox{H}^+ + \mbox{CN}(j').
\label{eq:proc1}
\end{equation}
The interaction of H$^+$ + CN is treated as a rigid rotor model. We used close-coupling quantum-mechanical methods to calculate the cross sections and collision rates of a CN molecule
with a proton  H$^+$.  The Schr\"odinger equation for an $a+bc$ collision in the center of 
mass frame, where $a$ (H$^+$ ) is an atom and 
$bc$ (CN) is a linear rigid rotor, is:
\cite{arthurs,green}
\begin{equation}
\left(\frac{P_{\vec R}{^2}}{2{M_R}} + \frac{L_{\hat 
r}{^2}}{2\mu r^2} + V(\vec r,\vec R) - E \right)
\Psi(\hat r,\vec R)=0.
\label{eq:schred}
\end{equation}
where $P_{\vec R}$ is the relative momentum between $a$ and $bc$,  
{${M}_R$} is the  reduced
mass of the atom-molecule (rigid rotor in this model) system $a+bc$:
${M_R} = m_a(m_b+m_c)/(m_a+m_b+m_c)$,
$\mu$ is the  reduced mass of the target:
$\mu=m_bm_c/(m_b + m_c)$,
$\hat r$ is the angle of orientation of the rotor $ab$, 
$V(\vec r, \vec R)$ is the PES for the 
three-atom system $abc$, and
$E$ is the total energy of the system.

To solve Eq.(2), the following 3-atom wave expansions are used \cite{green2}:
\begin{equation}
\Psi(\hat r,\vec R)=\sum_{JMjL}\frac{U^{JM}_{jL}(R)}{R}
\phi^{JM}_{jL}(\hat r,\vec R),
\label{eq:expan}
\end{equation}
and
\begin{eqnarray}
\phi^{JM}_{jL}(\hat r,\vec R) = \sum_{m_1m_2}
C_{jm_1Lm_2}^{JM}  Y_{jm_1}(\hat r)  Y_{Lm_2}(\hat R),
\end{eqnarray}

A set of coupled differential equations are achieved by solving the Schršdinger equation through the use of
the free software MOLSCAT (ver.14, J. Hudson and S. Green (1994)): 
\begin{eqnarray}
&&\left(\frac{d^2}{dR^2}-\frac{L(L+1)}{R^2}+k_{jL}^2\right)U_{jL}^{JM}(R)
=2 {M_R}
\sum_{j'L'} \int <\phi^{JM}_{jL}(\hat r,\vec R)  
|V(\vec r,\vec R)| 
\phi^{JM}_{j'L'}(\hat r,\vec R)> \nonumber \\
&&\times U_{j'L'}^{JM}(R) d\hat r d\hat R.
\label{eq:cpld}
\end{eqnarray}

The formulas used for analyzing the cross sections are  related to the 
asymptotic behavior of the H$^+$  + CN wave function using
$U^{JM}_{jL}(R\rightarrow\infty)$. 
The numerical results are fitted to the known asymptotic behavior 
of $U^{JM}_{jL}(R)$ from \cite{landau}:

\begin{eqnarray}
U_{jL}^{JM}
&\mathop{\mbox{\large$\sim$}}\limits_{R \rightarrow + \infty}&
\delta_{j j'} \delta_{L L'}
e^{-i(k_{\alpha}R-(L\pi/2))} 
 - \left(\frac{k_{\alpha}}{k_{\alpha'}}\right)^{1/2}S^J(j'L';jL;E) 
e^{i(k_{\alpha'}R-(L'\pi/2))},
\end{eqnarray}
and

\begin{eqnarray}
\sigma(j',j,\epsilon)&=&\frac{\pi}{(2j+1)k^2_{\alpha}}
\sum_{JLL'}(2J+1)|\delta_{jj'}\delta_{LL'} -            
S^J(j'L';jL; E)|^2.
\label{eq:cross}
\end{eqnarray}
The kinetic energy is 
$\epsilon=E-B_ej(j+1)$,
where $B_e$ is the rotation constant of the rigid rotor $bc$, i.e. the hydrogen molecule.

\subsection{CN + H$^+$ Interaction Potential}
The PES equations used were found from works \cite{jamieson, crawford, depristo}. Where U$_{\textrm{p}}$ is the polarization potential, U$_{\textrm{d}}$  is the proton-dipole potential, and U$_{\textrm{q}}$ is the proton-quadrupole potential. The values for the potential parameters for these equations were found and referenced in \cite{crawford, crawford2, allison}. 

\begin{equation}
U_{P+CN}(\vec r) = U_p(\vec r)  + U_d(\vec r) + U_q(\vec r)
\end{equation}


\begin{eqnarray}
U_p(\vec r) = \left\{
\begin{array}{l}
-e^2\alpha_0\cdot (r^2+r_0^2)^{-2}/2,\   \mbox{if}\  r<r_s  \nonumber \\
-e^2\alpha_0/2 (r^2+r_0^2)^2 - [e^2\alpha_0\cdot (r-r_s)^2\cdot P_2(cos(\theta))]/ [2 (r^2+r_0^2)^2 \\ \nonumber
(b^2+(r-r_s)^2)], \ \mbox{if}\   r \ge r_s,
\end{array}\right.
\end{eqnarray}
\begin{eqnarray}
U_d(\vec r) = \left\{
\begin{array}{l}
0 ,\   \mbox{if}\  r \leq r_s  \nonumber \\
(eD\cdot (r-r_s)^2 P_1(cos(\theta)))/(r^2\cdot[b^2+(r-r_s)^2]), \mbox{if}\   r > r_s,
\end{array}\right.
\end{eqnarray}
\begin{eqnarray}
U_q(\vec r) = \left\{
\begin{array}{l}
0 ,\   \mbox{if}\  r < r_s  \nonumber \\
eQ / r^3 \cdot (r-r_q)^2 P_2(cos(\theta)) (b^2+(r-r_q)^2)^{-1}, \mbox{if}\   r \geq r_s,
\end{array}\right.
\end{eqnarray}

Where r is the vector that joins the center of mass of the CN molecule to the H$^+$, and ${\theta}$  is the angle between the r and the axis of the CN molecule. The potential surface is modeled in Fig. 1.

\section{RESULT and  DISCUSSION}
We have taken test calculations to insure that the
convergence of the results with respect to all parameters that enter
in the propagation of the Schr\"odinger equation exists. The parameters include the
atomic-molecular distance $R$, the total angular momentum $J$,
the number of total rotational levels to be included in the close-coupling
expansion and others.
Attention has been given to the total number of numerical
steps in the propagation over the distance $R$ of the Schr\"odinger Eq.
(\ref{eq:cpld}). The parameter $R$ ranges from 0.75 a.u. to 20.0 a.u.
Up to 50000 propagation points have been used as well as different mathematical propagation schemes included in MOLSCAT.

Fig.\ 2 shows our results for the rotational energy transfer cross-sections of the process in Eq. (\ref{eq:proc1}) when j transitions from 0 to 5. The results converge better at higher total energy. Fig.\ 3 shows separately the results for j transitions from 0 to 1 as compared to work [1]. We can see from Fig. \ 3 that the pattern of the graph obtained from our results is similar to work [1], but not exact. In Fig.\ 4 and Fig. \ 5, we compare different J values from our data. From the graphs, we can see that the cross sections are close. 

In summary, the subject of this work is the quantum-mechanical study of the state-resolved rotational relaxation and excitation cross sections in ultra-cold collisions 
between CN molecules and protons H$^+$.  We analyzed the excitation of the CN molecule. A model 3-body PES for H$^+$-CN has been adopted from a previous work \cite{jamieson} and the potential  is  shown in Fig.\ 1. 
Calculation for total elastic scattering cross sections
and for low quantum rotational transition states have been performed.
A test of the numerical convergence was taken. 

In the future, we plan to compute the  thermal rate coefficients.

\clearpage
\begin{figure}
\begin{center}
\includegraphics*[scale=1.0,width=45pc,height=30pc]{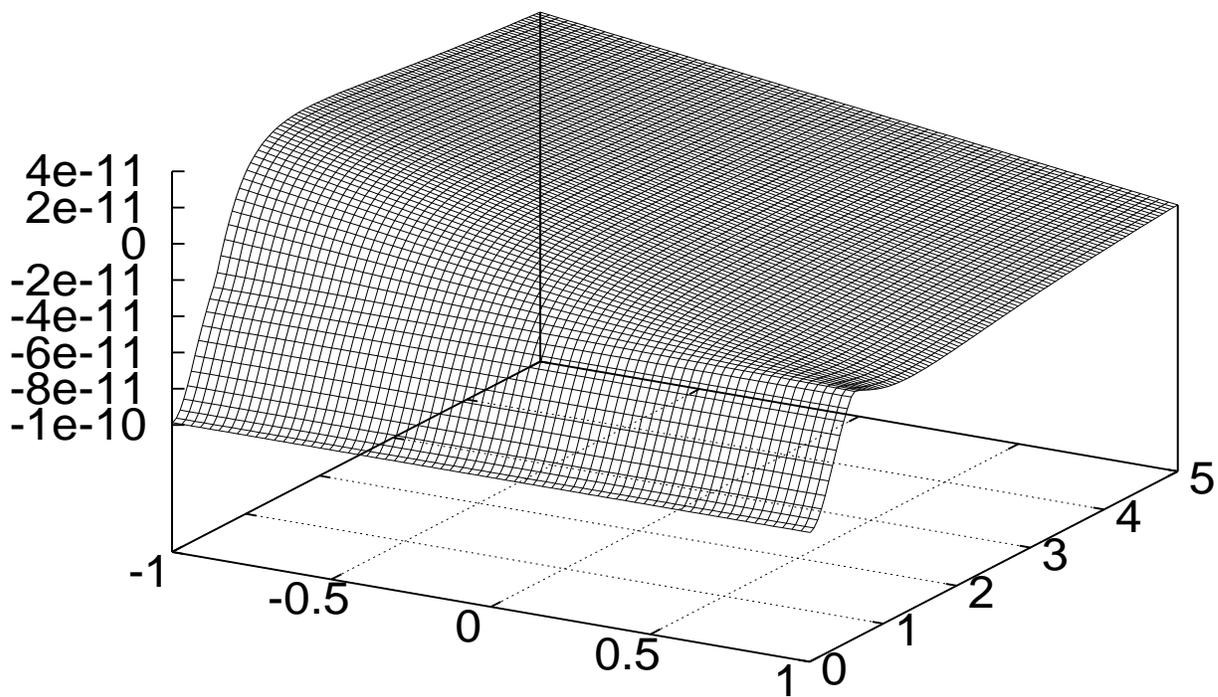}
\end{center}
\caption{The H$^+$ + CN potential energy surface used in this work.
}\label{fig1}
\end{figure}

\clearpage

\begin{figure}
\begin{center}
\includegraphics*[scale=1.0,width=37pc,height=27pc]{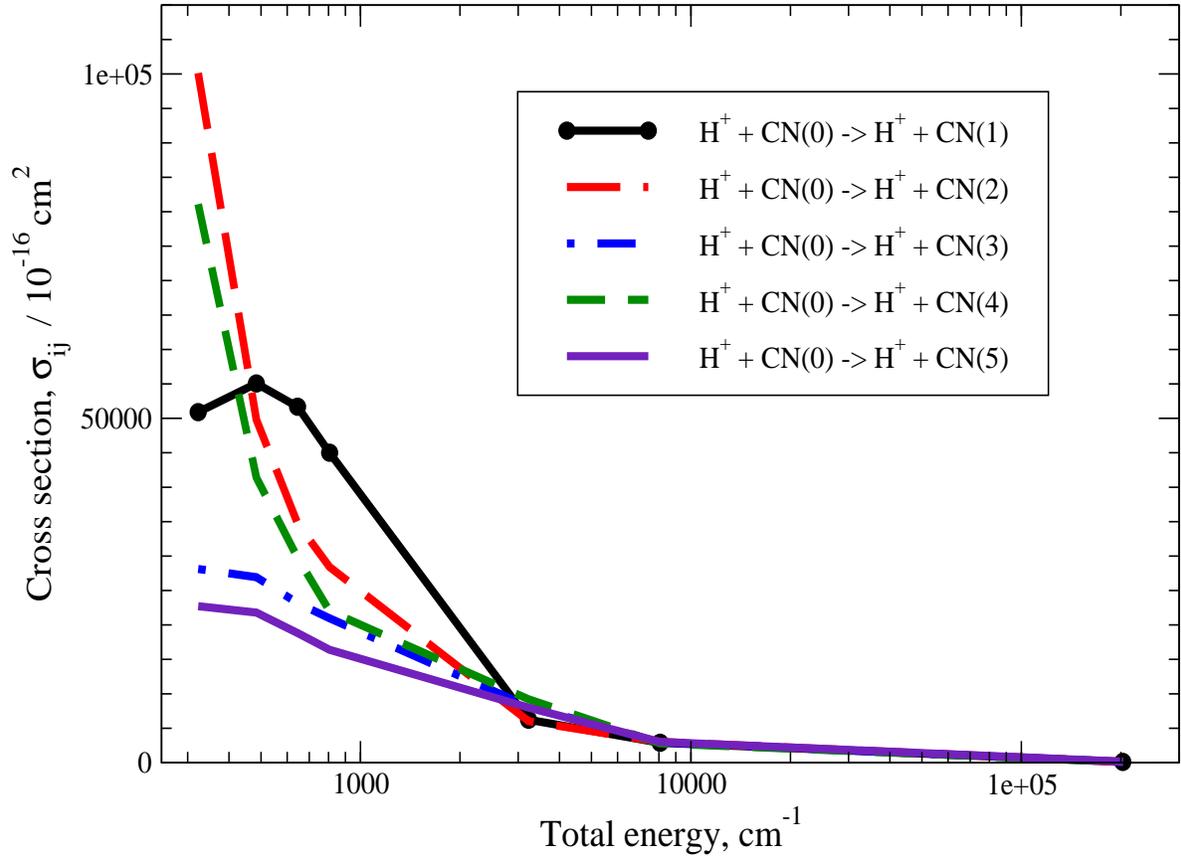}
\end{center}
\caption{The H$^+$ + CN Rotational Energy Transfer Cross-sections 
}\label{fig1}
\end{figure}

\clearpage

\begin{figure}
\begin{center}
\includegraphics*[scale=1.0,width=37pc,height=27pc]{B_7-7_1-2.eps}
\end{center}
\caption{Results for the  H$^+$  + CN (0) to H$^+$  + CN(1) 
}\label{fig1}
\end{figure}

\clearpage

\begin{figure}
\begin{center}
\includegraphics*[scale=1.0,width=37pc,height=27pc]{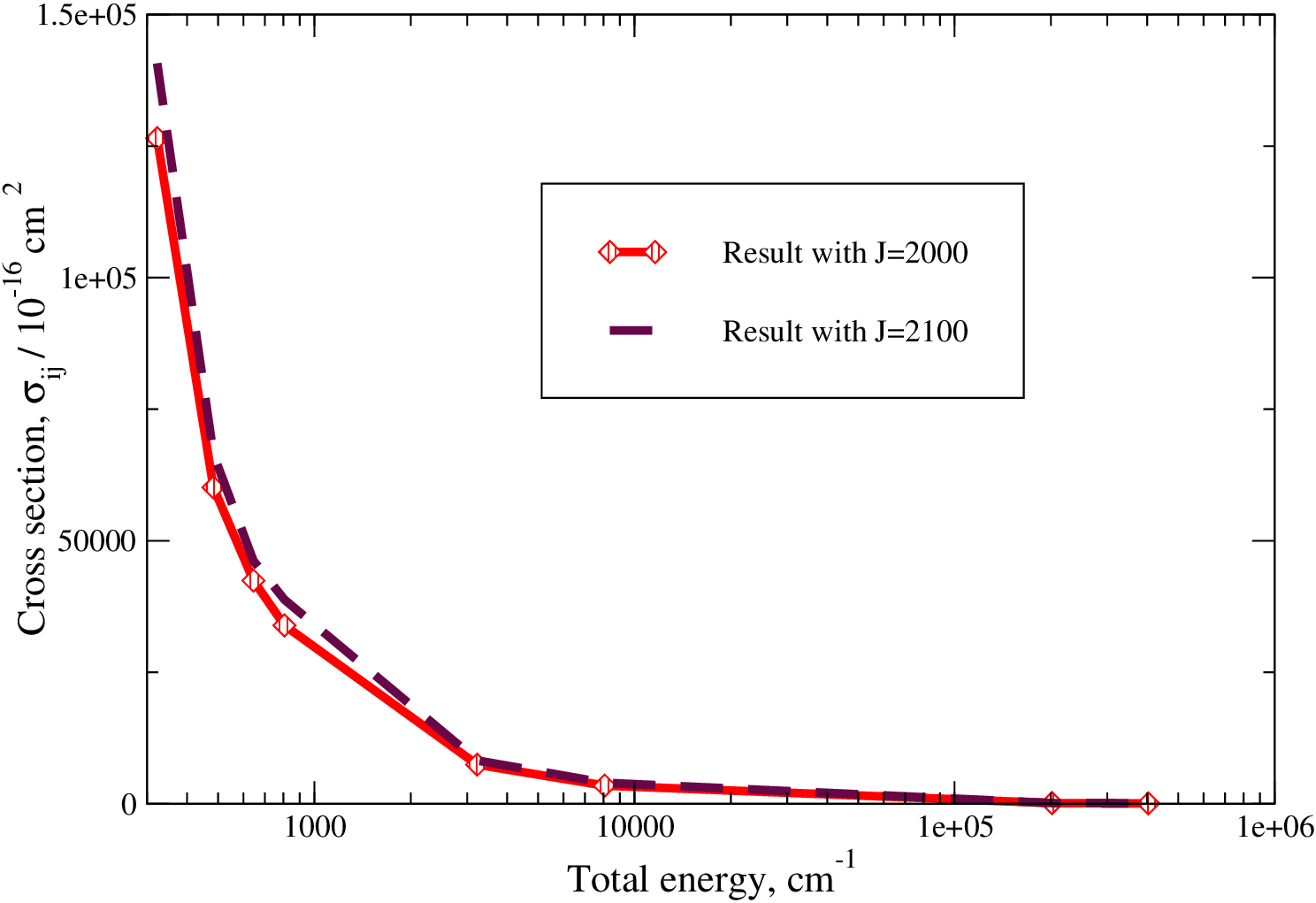}
\end{center}
\caption{Results for the  H$^+$  + CN (0) to H$^+$  + CN(2) 
}\label{fig1}
\end{figure}

\clearpage

\begin{figure}
\begin{center}
\includegraphics*[scale=1.0,width=37pc,height=27pc]{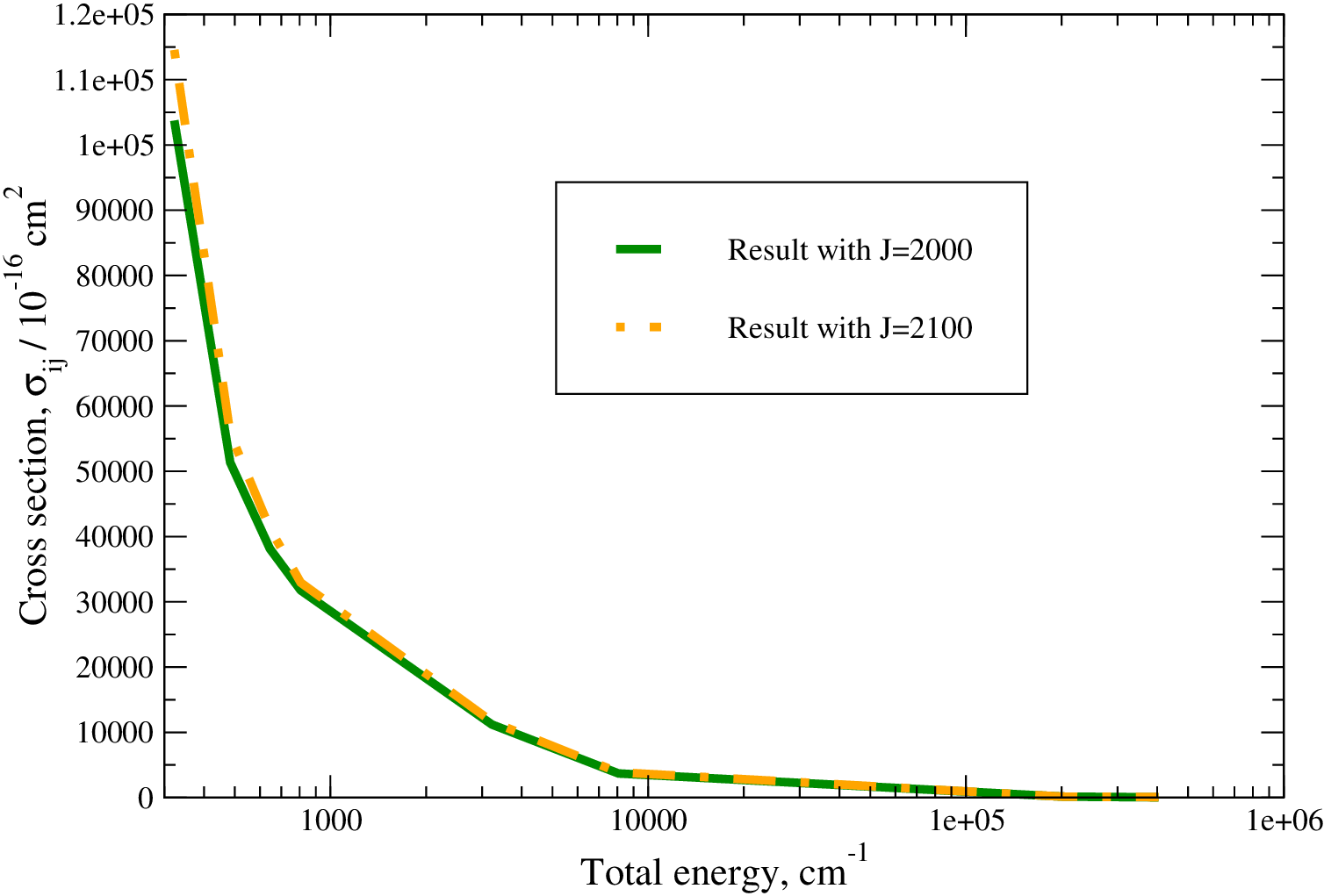}
\end{center}
\caption{Results for the  H$^+$  + CN (0) to H$^+$  + CN(4) 
}\label{fig1}
\end{figure}

\end{document}